# Coexistence of Paramagnetic-Charge-Ordered and Ferromagnetic-Metallic Phases in La$_{0.5}$Ca$_{0.5}$MnO$_3$ evidenced by ESR

F. Rivadulla[*], M. Freita-Alvite, and M. A. López-Quintela.
*Departamento de Química-Física, Universidad de Santiago de Compostela, E-15782 Santiago de Compostela, Spain.*

L. E. Hueso, D. R. Miguéns, P. Sande, and J. Rivas.
*Departamento de Física Aplicada, Universidad de Santiago de Compostela, E-15782 Santiago de Compostela, Spain.*

**Abstract.** Throughout a complete Electron Spin Resonance (ESR) and magnetization study of La$_{0.5}$Ca$_{0.5}$MnO$_3$, we discuss about the nature of the complex phase-segregated state established in this compound below T~210 K. Between T$_N$≤T≤T$_C$, the ESR spectra shows two lines characteristic of two different magnetic phases. From the resonance field ($H_r$) derived for each line we argue that the incommensurate-charge-ordering phase (ICO) which coexists with ferromagnetic-metallic (FMM) clusters in this temperature interval, is mainly paramagnetic and not antiferromagnetic. The FMM/ICO ratio can be tuned with a relatively small field, which suggests that the internal energy associated with those phases is very similar.

Below T$_N$, there is an appreciable FM contribution to the magnetization and the ESR spectra indicates the presence of FM clusters in an antiferromagnetic matrix (canted). Our results show that ESR could be a very useful tool to investigate the nature of the phase-separated state now believed to play a fundamental role in the physics of mixed valent manganites.

[*] corresponding author e-mail: qffran@usc.es

Since the possibility of an intrinsically phase-separated (PS) ground state in mixed-valent manganites was suggested by theoretical studies,[1,2,3] much experimental effort has gone into determine the exact nature of this inhomogeneous state.[4,5,6,7] As a consequence, the majority of works, early focused around the optimum doping level for $T_C$ of x~3/8, are now moving toward x~1/2, where the competition between these phases can be better studied.[8,9,10] At the 1:1 $Mn^{3+}$:$Mn^{4+}$ composition, charge ordering (CO), that inhibits the electron transfer associated with Double Exchange-Ferromagnetism (DE-FM) in manganites[11], is particularly feasible, leading to a rich variety of charge/orbital ordered (CO/OO) structures.[12] For example, previous studies showed that, upon cooling, $La_{0.5}Ca_{0.5}MnO_3$ becomes first ferromagnetic (FM) at $T_C$~225 K and then antiferromagnetic (AF) at $T_N$~155 K (180 K upon warming).[13] Its low temperature ground state shows a complex CO/OO-AF (CE-type) in which only anisotropic superexchange interactions, associated with Jahn-Teller (JT) distorted $Mn^{3+}O_6$ octahedra, are active.[14,15] Couplings along the [001] direction are AF, while in the (001) plane $Mn^{3+}$-O-$Mn^{4+}$ superexchange is FM (AF) when the occupied (unoccupied) $e_g$ orbitals are directed toward empty $Mn^{4+}$-$e_g$ orbitals.[14] Neutron powder diffraction revealed two different coherence lengths for the two interpenetrating $Mn^{3+}$-$Mn^{4+}$ magnetic sublattices in the CE structure, which implies the presence of structural/magnetic domain boundaries associated with the CO domain boundaries observed by electron diffraction.[15] On the other hand, x-ray synchrotron[15] and electron diffraction[16] experiments showed that the first order-AF transition at ~150 K is associated to an incommensurate-to-commensurate-CO transition (ICO-to-CCO), and that ICO and FM do coexist between 160 K≤T≤210 K as a finite spatial mixture of two competing phases (mutually exclusive). Moreover, the simultaneous presence of multiple phases (their relative proportion varying with temperature) with different degrees of orientational order of the JT distorted $Mn^{3+}O_6$ octahedra, has been demonstrated in this transition region.[8,12]

All these experiments show the extreme complexity of the magnetic/orbital structures in $La_{0.5}Ca_{0.5}MnO_3$, and demonstrates that much more work is still needed for a deep known of the mixed-phase state in manganites.

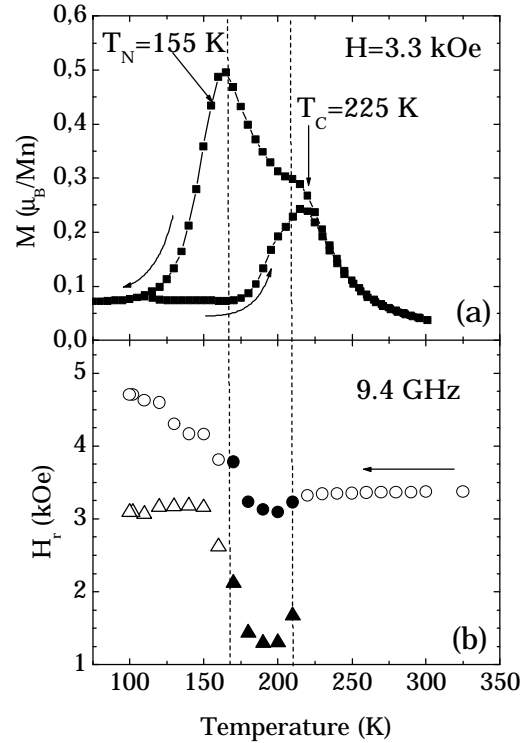

Figure 1: (a) Thermal evolution of the magnetization obtained in warming (after zero field cooling) and cooling. Vertical dashed lines are guides to the eye. (b) X-Band resonance fields ($H_r$) for the different lines (recording the ESR spectra on cooling) are plotted against temperature for comparison with the M(T) data.

To investigate the nature of the magnetic phases in $La_{0.5}Ca_{0.5}MnO_3$ at different temperatures, we carried out a detailed ESR study in a broad temperature interval. The ESR technique is very sensitive to magnetic heterogeneity, even if it involves a relatively small number of spins. Here we argue that the ICO is associated with a PM structure, which coexists with FMM regions. The ratio FMM/ICO-PM is strongly field dependent, even for relatively small fields (~0.5T).

Ceramic samples used for this study were synthesized by solid state reaction. Room temperature x-ray diffraction patterns indicate that the samples are single phase (orthorrombic, *Pnma*). Lattice parameters derived from Rietveld analysis (*a*=5.4241(5) Å, *b*=7.6479(1) Å, *c*=5.4353(1) Å) are in perfect agreement with available data.[15] ESR measurements were performed at 9.4 GHz (X-Band) with a EMX Bruker spectrometer between 100 K and 400 K. A small quatity of sample (~1 mg) was used for ESR experiments in order to avoid overloading of the cavity. Magnetization (versus temperature and field) was measured in a SQUID magnetometer.

In figure 1a) we show the thermal evolution of the magnetization measured on warming and cooling at 3.3 kOe (the resonance field for a free electron at X-Band, in order to compare it with ESR data). The strong thermal hysteresis (~30 K) resembles the first-order character of the transition. On the other hand, a sudden increase of resistivity (not shown) is also observed at $T_N$, indicating the appearance of CO associated to the AF structure.[16]

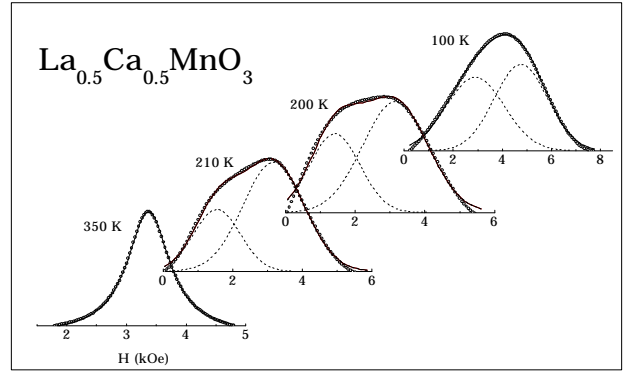

Figure 3: Integrated ESR spectra at various temperatures along with the fit to multiple lines.

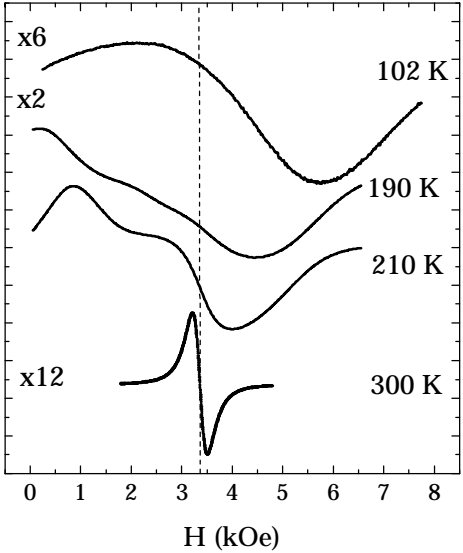

Figure 2: X-Band ESR spectra at different temperatures. Between $T_C$ (~225 K) and $T_N$ (~255 K) two lines are clearly observable.

Above $T_C$ the ESR spectra consist of a single lorentzian line centered at $g$=2.00 independent of temperature (Fig. 1b) and Fig. 2). As a function of T, both ESR linewidth and intensity showed a behavior similar to that described for other manganites in the paramagnetic state: the linewidth decreases with temperature on approaching $T_C$ from above, goes through a minimum at ~$1.1T_C$ and increases again.[17,18] The intensity follows a Curie-Weiss law in this regime, *i.e.* all the spins contribute to ESR.

In figure 2 we shown some ESR lines, representative of the general behavior in the different temperature intervals. When reducing temperature below ~215 K, the spectra split in two lines (low field, LF; and high field, HF, lines) that are observable down to ~160K where the long-range AF develops ($T_N$).

Below this temperature a single (apparently) broad line is recovered, with a significant increment of the resonance field. When the ESR spectrum is recorded increasing temperature, the broad line characteristic of the low temperature AF phase is maintained up to ~170 K, following the hysteretic behavior observed in M(T).

We have fitted the ESR spectra keeping the resonance field ($H_r$), the linewidth, and the intensity of each line (when more than one line is employed in the fitting procedure) as adjustable parameters. Some of these fits are shown in figure 3. Above $T_C$, the ESR spectra can be successfully reproduced with a single absorption line (lorentzian). On the other hand, as we mentioned above, between $T_N$ and $T_C$ the spectra splits in two lines, (LF, HF) characteristic of two magnetic components, their relative proportions varying with temperature. Below $T_N$, a single line is recovered, but only apparently, as again two lines are needed to reproduce the experimental data. When three or more lines were employed to fit the ESR the spectra, the goodness of fit does not improves, demonstrating the validity of our fits with two lines in this temperature range. Only between 150 K-160 K, the presence of three different phases cannot be discarded from our data. In fact, Radaelli *et al.*[8] suggested the possibility of four different phases corresponding to JT-distorted $MnO_6$ octahedra with different orientations of the *$d_{z^2}$* orbital, as we mentioned in the introduction of the paper.

The $H_r$ lines derived from the fits are shown in figure 1b) for better comparison with M(T) data. In polycrystalline samples, resonance data can be fitted to the equation

$$w = g H_r \qquad [1]$$

where the resonance field $H_r$ is the sum of the applied ($H_a$) and effective internal field ($H_i$), and $g=ge/2mc$.[19] For a system of randomly oriented and non interacting spherical crystalites of a ferromagnetic materials with cubic anisotropy, a first order approximation yields $H_r=H_a-H_K$, where $H_K=K_1/M_S$ represents the anisotropy field ($K_1$ is the anisotropy constant and $M_S$ the saturation magnetization).[19] In the ferromagnetic region the internal field adds to the applied field the resonance condition for a fixed frequency is reached at lower values of $H_a$. On the other hand, for a pure PM material $H_i=0$ and $H_a=H_r$ (3.3 kOe at 9.4 GHz). Moreover, for a purely AF material the ESR should not be observed at X-Band, as will be discussed later. All these facts make possible to assign the LF/HF lines observed in $T_N \leq T \leq T_C$ to FM/PM phases respectively.

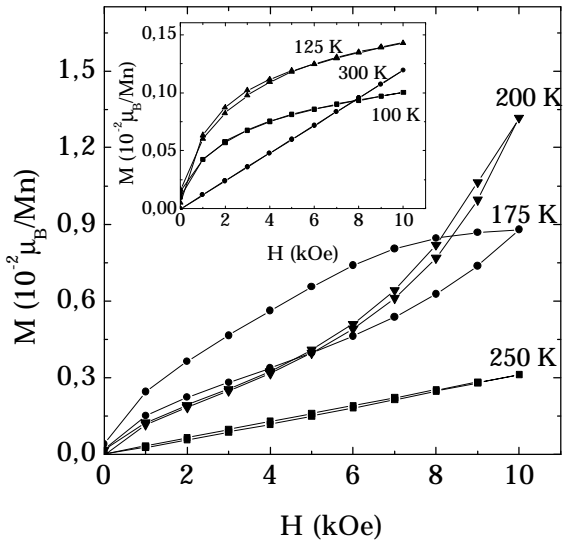

Figure 4: Field dependence of the magnetization of $La_{0.5}Ca_{0.5}MnO_3$ between $T_C$ and $T_N$ (main panel). There is a critical field around 0.5T which produces a sudden increase of magnetization. Inset: well above $T_C$ the expected linear (PM) behavior is observed, while the curves below $T_N$ indicates the presence of a net magnetic moment in the AF phase.

This is one of the key findings of our paper: the ICO state that coexists with the FMM regions between $T_C$ and $T_N$ is mainly PM and not AF. Preivous studies by $^{139}La$ NMR and ESR in $La_{0.5}Ca_{0.5}MnO_3$ suggested the coexistence of FM/AF phases between $T_N$ and $T_C$.[20] So, our results provide the first experimental evidence of the PM nature of the incommensurate CO phase in $La_{0.5}Ca_{0.5}MnO_3$. The value of $H_r$ ~3150 Oe for the HF line indicates a small internal field from the FM regions acting on the CO-PM phase.

The fraction of the FM phase can be increased by the effect of applied external field. In figure 4 magnetization isotherms are shown at different temperatures. Between $T_C$ and $T_N$, the slope of the M(H) curves increases at a critical field of ~0.5T. The sudden increase in the magnetization marks the increasing volume fraction of the FM phase as the ICO-PM phase is melted by the applied field. This effect increases the number of free carriers (via itinerant de Gennes double-exchange[21]) and hence produces a diminution of the resistivity. To check this hypothesis, we measured the magnetoresistance at different temperatures (figure 5). It is clear from the large values obtained between $T_C$ and $T_N$ that the metallic fraction is nicely increased by an small applied field in this temperature interval, while only small values of MR are reached below $T_N$. In fact, Roy et al.[22] previously found an appreciable MR for $H \geq$~0.5T in $La_{1-x}Ca_xMnO_3$ with %$Mn^{4+}$~52.8%-53.2%. Iodometric analysis in our samples gave $Mn^{4+}$~51.4%, which coincides satisfactorily well with their results.

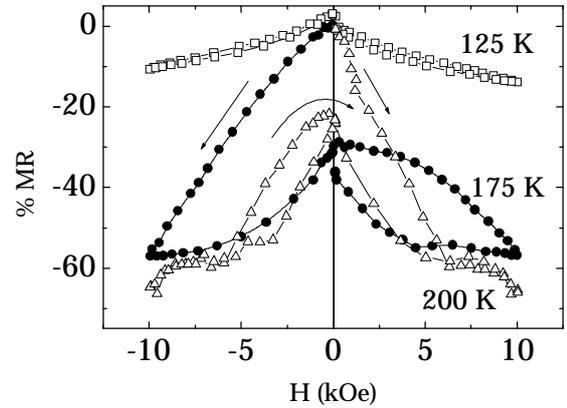

Figure 5: Magnetoresistance at different temperatures. Note the large values of MR obtained between $T_C$ and $T_N$, where the effect of the applied field over magnetization is stronger. The large hysteresis and the anomalies in the R(H) curves reproduce the M(H) behavior.

Extrapolating to H=0 from the initial magnetization curves before the critical field and comparing with the theoretical FM value of 3.5 $\mu_B$ for $Mn^{3+}/Mn^{4+}=1$, we have obtained the fraction of the FMM phase at H=0 and its temperature dependence (Figure 6). From the values of M(H=1T) it follows that the percentage of FM phase increases dramatically with applied field between $T_C$ and $T_N$, leading to the large values of MR at relatively small fields here presented.

The relative fraction of the FM phase extracted from ESR experiments is also proportional (no absolute values were obtained) to those from M(H), at least between $T_C$ and $T_N$.

It is clear that even in the low temperature AF state, there is a FM contribution to the magnetization which is about 3% of the full FM moment at H=0. To elucidate whether this ferromagnetism comes from a canted AF state or from small metallic clusters embedded in the AF matrix we have studied the ESR spectra at $T<T_N$.

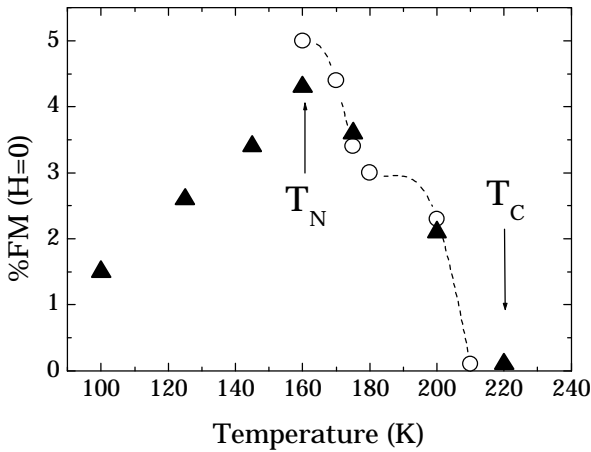

Figure 6: Experimental percentage of the FM phase obtained from M(H) (▲). Data from ESR experiments (○) are also plotted to shown the proportionality with magnetic data, although we could not get absolute values from the ESR experiments.

Below $T_N$, the broad ESR lines can be decomposed as the sum of two independent lines (see Fig. 3) one at $H_r$~3100 Oe and other that varies from $H_r$~4150 Oe at 150 K to $H_r$~4750 Oe at 100 K (Fig. 1b)). The line at low field indicates the presence of FM clusters even at these low temperatures, in a similar configuration as it was recently suggested for $La_{0.35}Ca_{0.65}MnO_3$ by $^{55}Mn$ and $^{139}La$ NMR.[7] On the other hand, as we have indicated in the introduction of this paper, the low temperature ground state in $La_{0.5}Ca_{0.5}MnO_3$ is AF. Normally, AF resonance is observed at very high frequencies, around 100 GHz or more, far beyond the resonance frequencies used in this experiment, and hence no line should be observed at 9.4 GHz. However, in the *Pnma* structure the oxygen atoms that mediate the interaction between Mn ions do not occupy inversion symmetry centers of the crystal, due to the canting of $MnO_6$ octahedra. In this situation antisymmetric superexchange interaction (Dzialoshinsky-Moriya coupling) between Mn produces spin canting and the appearance of a weak ferromagnetic moment. In this situation, two non-degenerate resonance modes appear, one at very high frequencies (not observed in our experiments at 9.4 GHz) while the other one will occur at ordinary microwave frequencies, and can be considered similar to a ferromagnetic mode.[19] For this reason, the resonance observed at high fields below $T_N$ can be attributed to the weak ferromagnetic component. The fact that the line is observable in the AF state is an indication of the existence of canting between antiparallel sublattices. Let us now compare our experimental findings with recent theoretical results about phase separation in x=0.5 manganites. Yunoki *et al.*[23] obtained the magnetic phase diagram for x=0.5 as a function of electron-phonon coupling (λ) and $J_{AF}$ spin exchange. For intermediate values of λ, as it is the case for $La_{0.5}Ca_{0.5}MnO_3$ [24], their results indicate the of CO and FM phases in an intermediate temperature range (see figures 3b) and 3d) in reference [23]). Moreover, the CE-AF structure is developed at lower temperatures, when the CO is well established, then supporting our hypothesis of an intermediate CO-PM state between $T_C$ and $T_N$. A systematic study of a $A^{3+}_{0.5}B^{2+}_{0.5}MnO_3$ series covering a wide $<r_A>$ interval, and hence a $J_{AF}$ range, will be very useful to elucidate the nature of the phase separation mechanism in half doped manganites, as well as the different phases implicated.

In summary, we have demonstrated that the magnetic structure of the incommensurate CO state stabilized between $T_N$ and $T_C$ in $La_{0.5}Ca_{0.5}MnO_3$ is PM and not AF. This ICO state can be melted by an external applied field (~0.5T) that grows the FM regions and increases the number of free carriers, leading to considerable values of MR. On the other hand, ESR and M(H) at $T<T_N$ indicate that the low temperature AF state in $La_{0.5}Ca_{0.5}MnO_3$ is also inhomogeneous. Our results are in perfect agreement with recent theoretical estimations.

Finally, ESR is here presented as a useful tool to investigate the magnetic structure of the phase-separated state in manganites.

**Acknowledgments**
The authors want to acknowledge Dr. E. Dagotto, for discussion and critical reading of the manuscript. We also acknowledge Spanish DGCYT for financial support under project MAT98-0416.

[1] S. Yunoki *et al.* Phys. Rev. Lett. **80**, 845 (1998).
[2] A. Moreo, S. Yunoki, and E. Dagotto, Science **283**, 2034 (1999); and references therein.
[3] G. Varelogiannis, Phys. Rev. Lett. **85**, 4172 (2000).
[4] M. Uehara, S. Mori, C. H. Chen, and S.-W. Cheong, Nature **399**, 560 (1999).
[5] M. Fath *et al.*, Science **285**, 1540 (1999).
[6] R. H. Heffner *et al.*, Phys. Rev. Lett. **85**, 3285 (2000).
[7] Cz. Kapusta, P. C. Riedi, M. Sikora, and M. R. I barra, Phys. Rev. Lett. **84**, 4216 (2000).
[8] P. G. Radaelli *et al.*, Phys. Rev. Lett. **75**, 4488 (1995).
[9] C. N. R. Rao, and A. K. Cheetham, Science **276**, 911 (1997).
[10] C. Ritter *et al.*, Phys. Rev. B **61**, R9229 (2000).
[11] C. Zener, Phys. Rev. **82**, 403 (1951).
[12] S.-W. Cheong, and C. H. Chen, in *Colossal Magnetoresistance, Charge Ordering and Related Properties of Manganese Oxides,* Edited by C. N. R. Rao and B. Raveau. (World Scientific, Singapore, 1998).
[13] P. Schiffer, A. P. Ramirez, W. Bao, and S.-W. Cheong, Phys. Rev. Lett. **75**, 3336 (1995).
[14] J. B. Goodenough, Phys. Rev. **100**, 564 (1955).
[15] P. G. Radaelli, D. E. Cox, M. Marezio, and S.-W. Cheong, Phys. Rev. B **55**, 3015 (1997)
[16] C. H. Chen, and S.-W. Cheong, Phys. Rev. Lett. **76**, 4042 (1996).
[17] M. T. Causa *et al.*, Phys. Rev. B **58**, 3233 (1998).
[18] F. Rivadulla *et al.*, Phys. Rev. B **60**, 11922 (1999).
[19] A. H. Morrish, *The Physical Principles of Magnetism*, (John Willey & Sons, NY, 1965).
[20] G. Papavassiliou *et al.*, Phys. Rev. B **55**, 15000 (1997).
[21] P. G. de Gennes, Phys. Rev. **118**, 141 (1960).
[22] M. Roy, J. F. Mitchell, A. P. Ramirez, and P. Schieffer, J.Phys.: Condens Matter **11**, 4843 (1999); M. Roy *et al.*, Phys. Rev.B **58**, 5185 (1998).
[23] S. Yunoki T. Hotta, and E. Dagotto, Phys. Rev. Lett. **84**, 3714 (2000).
[24] From resistivity data, and with $\varepsilon_F \sim 1.5$ eV, we have derived a value of $\lambda \sim 3.4$ for $La_{0.5}Ca_{0.5}MnO_3$.